\documentclass[traditabstract]{aa}
\usepackage{graphicx}
\begin{document}

\authorrunning{Nagao et al.}
\titlerunning{Infrared Metallicity Diagnostics}

\title{Metallicity diagnostics with infrared fine-structure lines}

\author{
          Tohru Nagao            \inst{1, 2, 3},
          Roberto Maiolino       \inst{4},
          Alessandro Marconi     \inst{5}, \and
          Hideo Matsuhara        \inst{6}
}

\offprints{T. Nagao}

\institute{
           Graduate School of Science and Engineering,
           Ehime University, 2-5 Bunkyo-cho, Matsuyama 790-8577, Japan
           \email{tohru@cosmos.ehime-u.ac.jp}
           \and
           Research Center for Space and Cosmic Evolution,
           Ehime University, 2-5 Bunkyo-cho, Matsuyama 790-8577, Japan
           \and
           Optical and Infrared Astronomy Division,
           National Astronomical Observatory of Japan,
           2-21-1 Osawa, Mitaka 181-8588, Japan
           \and
           INAF -- Osservatorio Astrofisico di Roma,
           Via di Frascati 33, 00040 Monte Porzio Catone, Italy
           \and
           Dipartimento di Fisica e Astronomia,
           Universit\`{a} di Firenze, Largo E. Fermi 2, 
           50125 Firenze, Italy
           \and
           Institute of Space and Astronautical Science, 
           Japan Aerospace Exploration Agency, 
           3-1-1 Yoshinodai, Chuo-ku, Sagamihara 252-5210, Japan
          }

\date{Received ;  accepted }

\abstract{
  Although measuring the gas metallicity in galaxies at various 
  redshifts is crucial to constrain galaxy evolutionary scenarios, only 
  rest-frame optical emission lines have been generally used to 
  measure the metallicity. This has prevented us to accurately 
  measure the metallicity of dust-obscured galaxies, and accordingly 
  to understand the chemical evolution of dusty populations, such as 
  ultraluminous infrared galaxies. Here we propose diagnostics of the 
  gas metallicity based on infrared fine structure emission lines, which 
  are nearly unaffected by dust extinction even the most obscured 
  systems. Specifically, we focus on fine-structure  lines arising mostly 
  from H{\sc ii} regions, not in photo-dissociation regions, to minimize 
  the dependence and uncertainties of the metallicity diagnostics
  from various physical parameters. 
  Based on photoionization models, we show that the emission-line
  flux ratio of ([O{\sc iii}]51.80+[O{\sc iii}]88.33)/[N{\sc iii}]57.21 is an
  excellent tracer of the gas metallicity. The individual line ratios
  [O{\sc iii}]51.80/[N{\sc iii}]57.21 or [O{\sc iii}]88.33/[N{\sc iii}]57.21 
  can also be used as diagnostics of the metallicity, but they suffer a 
  stronger dependence on the gas density. The line ratios 
  [O{\sc iii}]88.33/[O{\sc iii}]51.80 and [N{\sc ii}]121.7/[N{\sc iii}]57.21 
  can be used to measure and, therefore, account for the dependences 
  on the of the gas density and ionization parameter, respectively.
  We show that these diagnostic fine-structure lines are detectable with 
  Herschel in luminous infrared galaxies out $z \sim 0.4$. Metallicity
  measurements with these fine-structure lines will be feasible 
  at relatively high redshift ($z\sim 1$ or more) with 
  SPICA, the future infrared space observatory.
  
}

\maketitle

\section{Introduction}

The metallicity of gas and stars in galaxies is one of the most 
important properties to distinguish various galaxy evolutionary 
scenarios, since metals result from the cumulative star-formation 
activity and gas inflow/outflow history in galaxies. Since the gas 
metallicity of galaxies is in most cases much easier to measure 
than the stellar metallicity, past studies on the metallicity of 
galaxies and its evolution have mostly focused on the metallicity 
of the gas phase (e.g., McCall et al. 1985; Zaritsky et al. 1994; 
Pettini et al. 2001). Several extensive and detailed studies have 
been performed on the gas metallicity in local galaxies, the 
connection with other galaxy properties and the metallicity 
evolution through the cosmic epochs (e.g. Tremonti et al. 2004; 
Savaglio et al. 2005; Erb et al. 2006; Maiolino et al. 2008; 
Mannucci et al. 2009, 2010). Such observational studies have 
provided strong constraints on evolutionary scenarios 
(e.g., de Rossi et al. 2007; Kobayashi et al. 2007; Brooks et al. 
2007; Finlator \& Dav\'{e} 2008).

The vast majority of previous studies on the gas metallicity in
galaxies used rest-frame optical diagnostics (see, e.g., 
Nagao et al. 2006a and references therein). However, such 
rest-frame optical diagnostics cannot probe the metallicity of the 
regions affected by significant dust extinction, which is the case 
for many star forming galaxies, especially at high redshift. 
This is because the fraction of dusty galaxies such as 
ultra-luminous infrared galaxies (ULIRGs) increases as a function
of redshift (e.g., Le Floc'h et al. 2005). Hence, 
optical metallicity studies probably probe only the outer, less 
extinguished region of star forming galaxies, which may deviate 
substantially from the metallicity of inner more active and more 
obscured regions of star formation. A clear indication of such a 
mismatch was recently found in high-z submillimeter galaxies 
by Santini et al. (2010), who used dust-mass measurements to 
infer an metals content an order of magnitude higher than inferred 
from optical metallicity diagnostics. If such mismatch applies to a 
significant fraction of actively star forming galaxies, this would 
imply a major revision of the past results on the metallicity 
evolution in galaxies based on optical diagnostics. Within this 
context we note that, even for starbursts which have been 
well-studied in optical, a combination of radio and infrared 
measurements has conclusively demonstrated that some of the 
most active star-forming sites are optically obscured (e.g., 
Gorjian et al. 2001; Vacca et al. 2002; Soifer et al. 2008).

The access to far-infrared diagnostics of the gas metallicity would
overcome the dust extinction problems plaguing optical 
measurements. Observational studies on gas metallicity exploiting 
infrared indicators have already been carried out by using Infrared 
Space Observatory ($ISO$). Verma et al. (2003) investigated 
mid-infrared spectra of 12 nearby starburst galaxies and found a 
strong correlation between gas metallicity and gas excitation, 
which is also seen in some optical studies (e.g., Nagao et al. 2006; 
Maier et al. 2006). Garnett et al. (2004) studied chemical properties 
of gas in M51 based on the $ISO$ data and found that the C/O 
abundance ratio in M51 is consistent with the solar neighborhood 
value. This infrared-based metallicity studies is expected to 
progress dramatically by using the on-going and forthcoming 
observational facilities such as Herschel, JWST, SPICA, and ALMA. 
Within this context, it is important to investigate in detail model 
predictions yielding calibrations between the gas metallicity and 
flux ratios of emission lines at long wavelengths. Motivated by this, 
we present the results of photoionization model calculations and 
the theoretical calibrations of gas metallicity diagnostics based on 
fine-structure emission lines in mid- and far-infrared wavelength 
ranges.

\section{The method}

\begin{figure}
\centering
\rotatebox{-90}{\includegraphics[width=8.5cm]{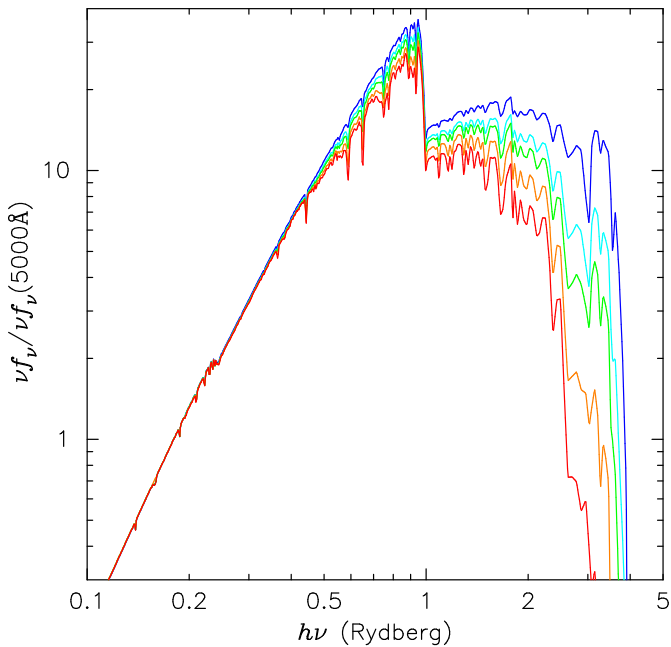}}
\caption{
   SEDs of the ionizing continuum adopted in our 
   photoionization models, which are
   starburst99 constant star-formation model spectra
   (Leitherer et al. 1999) with an age of 1 Myr and an 
   $\alpha = 2.35$ IMF ($M_{\rm low} = 1 M_\odot$ and 
   $M_{\rm up} = 100 M_\odot$).
   Although most of our analysis adopt these spectra, models 
   with different ages (3, 10, and 30 Myr) are investigated
   in \S3.3. 
   The blue, cyan, green, orange, and red spectra denote
   a model with $Z_{\rm star} = 0.05 Z_\odot$, $0.2 Z_\odot$,
   $0.4 Z_\odot$, $1.0 Z_\odot$, and $2.0 Z_\odot$,
   respectively. The $Z_{\rm star} = 2.0 Z_\odot$ spectrum 
   is used also for our $Z_{\rm gas} = 3.0 Z_\odot$ 
   calculations (see the main text).
   All spectra are normalized at 5000${\rm \AA}$
   ($\simeq$ 0.18 Ryd).
}
\label{fig01}
\end{figure}

In the mid-infrared to sub-millimeter wavelength range, there are 
many fine-structure emission lines radiated by various ions in 
various excitation levels. Although some of them arise within 
H{\sc ii} regions, some others arise in photo-dissociation regions 
(PDRs) that surrounds the inner H{\sc ii} regions. For instance the 
[C{\sc ii}]157.6 emission, that is the strongest emission line in 
spectra of most galaxies and thus detected even in very distant 
galaxies (e.g. Maiolino et al. 2005; Iono et al. 2006; 
Maiolino et al. 2009, Hailey-Dunsheath et al. 2010), is mainly from 
PDRs rather than H{\sc ii} regions (e.g., Crawford et al. 1985; 
Stacey et al. 1991). Some emission lines arise both in H{\sc ii} 
regions and PDRs (e.g., Petuchowski \& Bennett 1993; Heiles 
1994; Aannestad \& Emery 2003; Abel 2006). The latter class of 
lines are the most difficult to model, since they have a more 
complex dependence on the various physical parameters of the 
clouds, and therefore they are not ideal metallicity diagnostics. 
Even lines arising mostly from PDR are not optimal to be used as 
metallicity diagnostics, since modeling PDRs is very challenging, 
because involves complex physical mechanisms of the gas 
heating and of radiative transport. The modeling of H{\sc ii} 
regions is much more easy and generally more understood and 
tested; in these regions the dependence of the line emissivity 
from the various physical parameters is much simpler than in the 
case of PDRs. As a consequence, lines arising mostly from
H{\sc ii} regions are more suitable candidates to measure the gas
metallicity of the inter-stellar matter. Therefore, it is very important
to understand  the emissivity distribution of various fine-structure 
emission lines in gas clouds and, specifically, understand which 
ones arise mostly from H{\sc ii} regions and which are instead 
significantly contributed by PDRs. This is possible by using the 
publicly available photoionization model code 
$Cloudy$, as described by Abel et al. (2005, 2006).

We have carried out photoionization model calculations by using 
$Cloudy$ version 08.00 (Ferland et al. 1998; Ferland 2008). We 
assume pressure-equilibrium gas clouds with a plane-parallel 
geometry. As described later, the results are not significantly 
different if we assume constant-density clouds instead of 
pressure-equilibrium clouds, as far as only H{\sc ii} regions are 
considered (see \S\S3.2). Calculations are stopped at the depth 
where the fraction of the H$_2$ molecule reaches 50\% of the total
hydrogen, when we investigate lines arising both from H{\sc ii} 
regions and from PDRs. This depth is defined as the outer edge of 
PDRs, following Abel et al. (2005). On the other hand, calculations 
are stopped at the depth where the fraction of H$^+$ reaches 
down to 1\% of the total hydrogen, when the emission lines arising 
only from H{\sc ii} regions are investigated (\S\S3.2). We define
this depth as the outer edge of H{\sc ii} regions.

Some low-excitation fine-structure lines can arise beyond this outer 
edge of PDRs (i.e., fully molecular regions surrounding the inner 
PDRs; see, e.g., Abel et al. 2005). We do not investigate such 
fully-molecular regions since our main interest in this work is on 
emission-lines arising within H{\sc ii} regions, as described below. 

\begin{table}
\centering
\caption{Adopted solar elemental abundance ratios and 
         gas-phase depletion factors}
\label{table:01}
\begin{tabular}{l l c}
\hline\hline
\noalign{\smallskip}

element          &
$n/n_{\rm H}$    &
depletion factor \\

\noalign{\smallskip}
\hline 
\noalign{\smallskip}

H  & 1.00e+0   & 1.00 \\
He$^\mathrm{a}$ & 1.03e--1 & 1.00 \\
Li & 2.04e--9  & 0.16 \\
Be & 2.63e--11 & 0.60 \\
B  & 6.17e--10 & 0.13 \\
C  & 2.45e--4  & 0.40 \\
N$^\mathrm{a}$  & 8.51e--5 & 1.00 \\
O  & 4.90e--4  & 0.60 \\
F  & 3.02e--8  & 0.30 \\
Ne & 1.00e--4  & 1.00 \\
Na & 2.14e--6  & 0.20 \\
Mg & 3.47e--5  & 0.20 \\
Al & 2.95e--6  & 0.01 \\
Si & 3.47e--5  & 0.03 \\
P  & 3.20e--7  & 0.25 \\
S  & 1.84e--5  & 1.00 \\
Cl & 1.91e--7  & 0.40 \\
Ar & 2.51e--6  & 1.00 \\
K  & 1.32e--7  & 0.30 \\
Ca & 2.29e--6  & 1e--4 \\
Sc & 1.48e--9  & 5e--3 \\
Ti & 1.05e--7  & 8e--3 \\
V  & 1.00e--8  & 6e--3 \\
Cr & 4.68e--7  & 6e--3 \\
Mn & 2.88e--7  & 0.05 \\
Fe & 2.82e--5  & 0.01 \\
Co & 8.32e--8  & 0.01 \\
Ni & 1.78e--6  & 0.01 \\
Cu & 1.62e--8  & 0.10 \\
Zn & 3.98e--8  & 0.25 \\  

\noalign{\smallskip}
\hline
\end{tabular}
\begin{list}{}{}
\item[$^\mathrm{a}$]
  See the main text for details of the treatments for helium
  and nitrogen.
\end{list}
\end{table}

The free parameters in our calculations are
(1) the hydrogen density of a cloud ($n_{\rm H}$) at the irradiated 
    surface of a cloud;
(2) the ionization parameter ($U$), i.e., the ratio of the ionizing 
    photon density to the hydrogen density at the surface of a cloud;
(3) the elemental composition of the gas; and
(4) the spectral energy distribution (SED) of the photoionizing 
    continuum radiation.
We investigate plane-parallel gas clouds with gas densities 
$n_{\rm H} = 10^1 - 10^4$ cm$^{-3}$ at the illuminated face (note 
that the gas density varies as a function of the depth within gas 
clouds), and ionization parameters $U = 10^{-3.5} - 10^{-1.5}$. 
Here the ionization parameter is a dimensionless parameter 
defined as follows:
\begin{equation}
   U = \frac{\Phi({\rm H})}{cn_{\rm H}} 
     = \frac{1}{cn_{\rm H}} \int_{\nu_1}^\infty \frac{F_\nu}{h\nu} \ d\nu
\end{equation}
where $c$ is the speed of the light, $\Phi$(H) is the surface flux of 
hydrogen-ionizing photons, $F_\nu$ is the surface energy flux of 
the input radiation, and $\nu_1$ is the frequency of the Lyman edge 
($h\nu_1 =$ 13.6 eV).

The adopted solar elemental abundance ratios (in terms of number 
density, not mass) and gas-phase depletion factors are given for 
convenience in Table 1. The adopted depletion factors are derived
from the literature (e.g., Cowie \& Songaila 1986; Jenkins 1987; 
Cardelli et al. 1991; see Ferland 2006 for more details). However, 
there is no broad consensus on those factors and, moreover, they 
may depend on the gas density (e.g., Spitzer 1985). We do not
discuss this issue further in this work, and we simply adopt the
depletion factors given in Table 1. For non-solar metallicities we 
assume that both the dust model and the depletion factors are 
unchanged, but the dust abundance is assumed to scale linearly 
with the gas metallicity. All elements except nitrogen and helium are 
taken to be primary nucleosynthesis elements. For helium, we 
assume a primary nucleosynthesis component in addition to the 
primordial value, that is,
\begin{equation}
   \frac{n_{\rm He}}{n_{\rm H}} = 
   0.0737 + 0.0293 \ Z_{\rm gas}/Z_\odot
\end{equation}
(Groves et al. 2004). Nitrogen is assumed to be a secondary 
nucleosynthesis element above metallicities of 0.23 solar, and a 
primary nucleosynthesis element at lower metallicities, i.e.,
\begin{equation}
   \log \frac{n_{\rm N}}{n_{\rm H}} = 
      2 \ \log \frac{Z_{\rm gas}}{Z_\odot} - 4.070 \ \ \ 
        {\rm when} \ Z_{\rm gas} > 0.23 Z_\odot \\
\end{equation}
and
\begin{equation}
   \log \frac{n_{\rm N}}{n_{\rm H}} = 
      \log \frac{Z_{\rm gas}}{Z_\odot} - 4.709 \ \ \ 
        {\rm when} \ Z_{\rm gas} < 0.23 Z_\odot \\
\end{equation}
(Kewley \& Dopita 2002). Note that the behavior of nitrogen as a 
secondary element (Eq.~3) is observationally confirmed through 
the data of H{\sc ii} regions in galaxies (e.g., van Zee et al. 1998; 
Pilyugin et al. 2003). Chemical evolutionary models of galaxies also 
predict such a behavior for a wide range of parameters, including 
variations of the IMF (e.g., Hamann \& Ferland 1993; Chiappini et al. 
2003). The validity of Eq.~3 is sometimes questioned at very high 
metallicities, $Z_{\rm gas} \geq 5 Z_\odot$, which is however 
beyond the range studied in this work (see, e.g., Nagao et al. 2006b; 
Jiang et al. 2008). We investigate gas clouds with gas metallicities
$Z_{\rm gas}$ = 0.05, 0.2, 0.4, 1.0, 2.0, and 3.0 times solar metallicity. 
In the calculations, Orion-type graphite and silicate grains 
(Baldwin et al. 1991; Ferland 2006), as well as polycyclic aromatic 
hydrocarbon (PAH), are included.

\begin{figure*}
\centering
\rotatebox{-90}{\includegraphics[width=10.0cm]{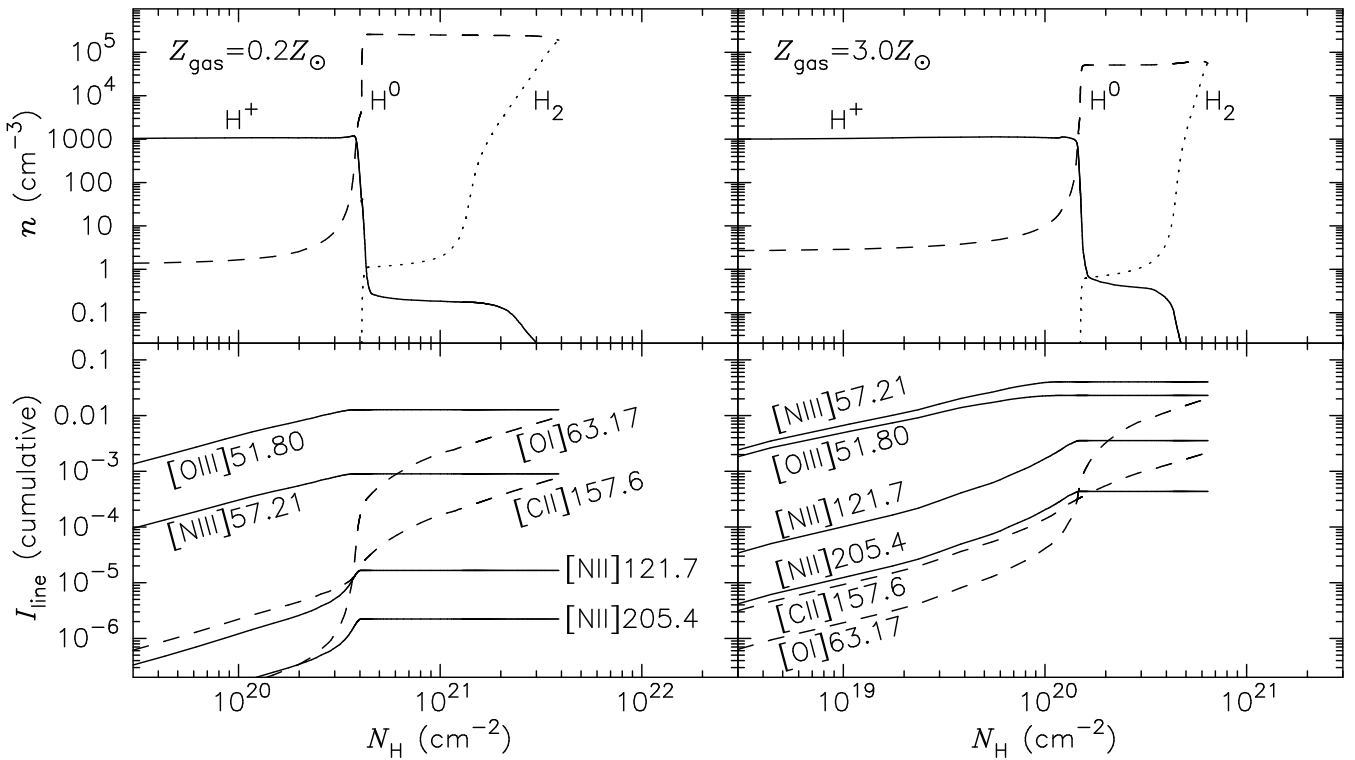}}
\caption{
   Hydrogen density structure (upper panels) and cumulative
   intensity of some fine-structure emission lines
   (lower panels) of clouds with $Z_{\rm gas} = 0.2 Z_\odot$
   (left hand panels) and $Z_{\rm gas} = 3.0 Z_\odot$ (right hand
   panels), as a function of hydrogen column density. For both
   cases, $n_{\rm H} = 10^3$ cm$^{-3}$ and $U = 10^{-2.5}$
   are adopted (where $n_{\rm H}$ is the hydrogen density at
   the illuminated face of clouds). Solid, dashed, and dotted 
   lines in the upper panels denote the densities of H$^+$, 
   H$^0$, and H$_2$, respectively. In the lower panels, solid
   lines show cumulative intensities for fine-structure 
   emissions mainly from H{\sc ii} regions while dashed lines 
   are for emissions that are emitted also, or mostly from PDRs.
}
\label{fig02}
\end{figure*}

The adopted ionizing continua are the starburst99 
constant star-formation model spectra (Leitherer et al. 1999)
with an age of 1 Myr and an $\alpha = 2.35$ IMF 
($M_{\rm low} = 1 M_\odot$ and $M_{\rm up} = 100 M_\odot$). We 
selected these model parameters because, in this work, we want to 
simulate actively star-forming galaxies (such as infrared galaxies 
and submillimeter galaxies). However, to demonstrate that our
results do not depend on the ionizing continuum shape, we will also 
investigate models adopting SEDs with older ages (\S3.3).
The starburst99 spectra are available for stellar metallicities of 
$Z_{\rm star}$ = 0.05, 0.2, 0.4, 1.0, and 2.0 times solar metallicity, 
which are shown in Figure 1. For models with 
$Z_{\rm gas} = 3.0 Z_\odot$, we adopt the starburst99 spectrum 
with $Z_{\rm star} = 2.0 Z_\odot$. The cosmic background emission 
from radio to X-ray is also included (see Ferland 2006 for details). 
However, this component does not affect the calculation results 
significantly. The heating and ionization by cosmic rays are also
included, which affects the thermal solutions especially in the PDRs 
(see, e.g., Ferland \& Mushotzky 1984; Ferland 2006).

\section{Results}

\subsection{Contributions from H~II regions and PDRs}

We first investigate from which parts of clouds emission lines originate
in order to explore optimal emission-line combinations
which can be used as good metallicity diagnostics. In Figure 2, we 
show hydrogen density structures and cumulative intensities of 
some fine-structure emission lines of clouds with 
$Z_{\rm gas} = 0.2 Z_\odot$ and $3.0 Z_\odot$, as functions of 
the hydrogen column density. The outer edges of PDRs defined 
above correspond to the end points of calculations. The outer 
edges of H{\sc ii} regions are clearly seen as a sharp drop of the 
density of H$^+$. It is shown in Figure 2 that relatively high 
excitation lines such as [O{\sc iii}]51.80 and [N{\sc iii}]57.21 arise 
within H{\sc ii} regions while lower excitation lines such as 
[O{\sc i}]63.17 and [C{\sc ii}]157.6 arise mostly in PDRs.

To see how H{\sc ii} regions and PDRs contribute to the intensities 
of various infrared fine-structure emission lines more quantitatively, 
we give the fraction of the H{\sc ii} region contribution to total 
emission-line intensity for the cases of 
$Z_{\rm gas} = 0.2 Z_\odot$ and $3.0 Z_\odot$, 
$n_{\rm H} = 10^{1.0}$ cm$^{-3}$ and $10^{3.0}$ cm$^{-3}$, 
and $U = 10^{-2.5}$ and $10^{-1.5}$, in Tables 2 and 3. The following 
fine-structure emission lines arise within H{\sc ii} regions without 
significant contribution from PDRs (at least in the parameter space 
investigated here):
[Ne{\sc ii}]12.81, [Ne{\sc iii}]15.55, [S{\sc iii}]18.67, 
[S{\sc iii}]33.47, [Ne{\sc iii}]36.01, [O{\sc iii}]51.80,
[N{\sc iii}]57.21, [O{\sc iii}]88.33, [N{\sc ii}]121.7, and 
[N{\sc ii}]205.4 (hereafter ``H{\sc ii} region lines'').
On the other hand, line intensities of the following low-ionization
emission lines are significantly contributed by the PDRs: 
[Si{\sc ii}]34.81, [O{\sc i}]63.17, [O{\sc i}]145.5, and 
[C{\sc ii}]157.6 (hereafter ``PDR lines''). Although the PDR lines 
arise mostly within PDRs, H{\sc ii} regions also contribute to PDR 
lines in some cases. Note that the contribution by H{\sc ii} regions 
given in Tables 2 and 3 for those low-ionization PDR lines are in 
some cases upper limits. This is because such fine-structure emission 
lines could be contributed also from molecular cloud regions that 
extend beyond PDRs (see, e.g., Abel et al. 2005).

It is evidently safe to avoid combinations of an H{\sc ii} region line 
and a PDR line to investigate possible candidate of 
metallicity-diagnostic flux ratios. We therefore focus only on 
H{\sc ii} region lines to avoid significant parameter dependences 
in emission-line flux ratios.

\subsection{Metallicity-sensitive emission-line flux ratios}

Metallicity diagnostics by using infrared fine-structure lines have 
been also discussed in some previous studies. The main approach 
consists in calculating abundances of ionic species directly based 
on their emissivity. The flux ($F_{{\rm X}^{+i}}$) of a given ionic 
species (X$^{+i}$) is the product of the density of the ion 
($n_{{\rm X}^{+i}}$), the electron density ($n_{\rm e}$), and the 
emissivity of the ion ($j_{{\rm X}^{+i}}$). Therefore, in H{\sc ii} 
regions (where $n_{\rm H^+} \simeq n_{\rm H}$ by definition), 
the following relation applies:
\begin{equation}
  \frac{F_{{\rm X}^{+i}}}{F_{{\rm H}^+}} =
    \frac{n_{{\rm X}^{+i}} \ n_{\rm e} \ j_{{\rm X}^{+i}}}
         {n_{{\rm H}^{+}}  \ n_{\rm e} \ j_{{\rm H}^{+}}} ,
\end{equation}
which can be re-written as
\begin{equation}
  \frac{n_{{\rm X}^{+i}}}{n_{\rm H}} =
    \left[\frac{F_{{\rm X}^{+i}}}{F_{{\rm H}^+}}\right]
    \left[\frac{j_{{\rm H}^{+}}}{j_{{\rm X}^{+i}}}\right]
\end{equation}
where the emissivities are calculated once the gas temperature is 
known from the photoionization model. Once ionic abundances are 
inferred by photoionization models, we can estimate the elemental 
abundances by applying appropriate ionization correction factors 
(ICFs). For instance, Verma et al. (2003) derived elemental 
abundances of Ne from [Ne{\sc ii}]12.81 and [Ne{\sc iii}]15.56, 
Ar from [Ar{\sc ii}]6.99 and [Ar{\sc iii}]8.99, and S from both
[S{\sc iii}]18.71 and [S{\sc iv}]10.51, by using hydrogen 
recombination lines such as Br$\alpha$, for nearby starburst 
galaxies based on Spitzer spectra (see also Wu et al. 2008). 
Panuzzo et al. (2003) also investigated the strategy to derive N 
abundance by using [N{\sc ii}]121.7, [N{\sc iii}]57.21, and Br$\alpha$.

However, the requirement of relying hydrogen recombination lines
makes this approach highly uncertain. Indeed, the relatively large 
wavelength separation between hydrogen recombination lines and 
metallic emission lines (especially for the case of N) is hard to be 
covered by the same instrument, and the use of different instruments 
or facilities may introduce significant systematic errors (due to, e.g., 
aperture effects and calibration errors) in the estimated metallicities. 
Moreover, hydrogen recombination lines are sometimes difficult due 
to the effect of stellar absorption features. In addition, the wavelength 
of Br$\alpha$ is not long enough to avoid significant dust extinction 
in many cases, since it is known that the amount of extinction toward 
the most obscured part of star formation in galaxies can reach up to 
$A_V \sim 100$ mag (e.g., Verma et al. 2003). Although Panuzzo et al. 
(2003) discussed the use of the radio continuum emission instead of 
hydrogen recombination lines to overcome the above difficulties, the 
radio continuum generally contributed by synchrotron emission, or 
even an active galactic nucleus, and therefore the relation between 
the radio continuum flux and the hydrogen line flux may be 
complicated. Therefore, ``good'' metallicity diagnostics should consist 
of fine-structure emission lines (not of hydrogen recombination lines) 
at long wavelengths, whose wavelength separation is not exceedingly 
large.

\begin{figure}
\centering
\rotatebox{-90}{\includegraphics[width=5.9cm]{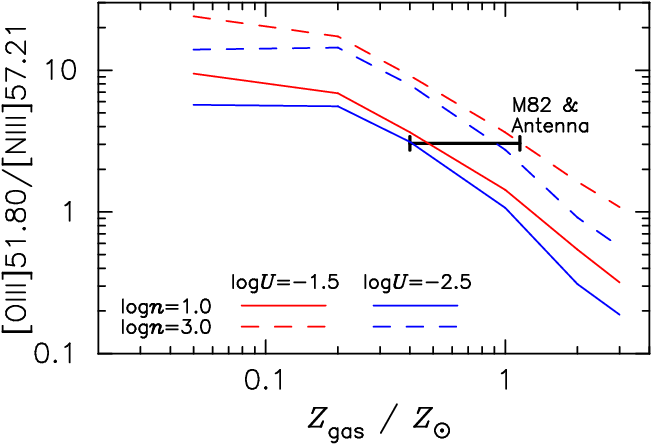}}
\caption{
   Predicted emission-line flux ratio of [O{\sc iii}]51.80/[N{\sc iii}]57.21 
   as a function of gas metallicity. Blue and red lines denote the 
   models with $U =$ 10$^{-2.5}$ and 10$^{-1.5}$, and solid and
   dashed lines denote the models with $n_{\rm H} =$
   10$^{1.0}$ cm$^{-3}$ and 10$^{3.0}$ cm$^{-3}$, respectively. The 
   flux ratio observed in M82 and in the Antennae galaxy is shown by 
   black horizontal line (these two galaxies show very similar 
   [O{\sc iii}]51.80/[N{\sc iii}]57.21 flux ratios; see Table 5).
   The $x$-range of this horizontal line corresponds to the inferred 
   metallicity range for M82 and the Antenna galaxy.
}
\label{fig03}
%
\rotatebox{-90}{\includegraphics[width=5.9cm]{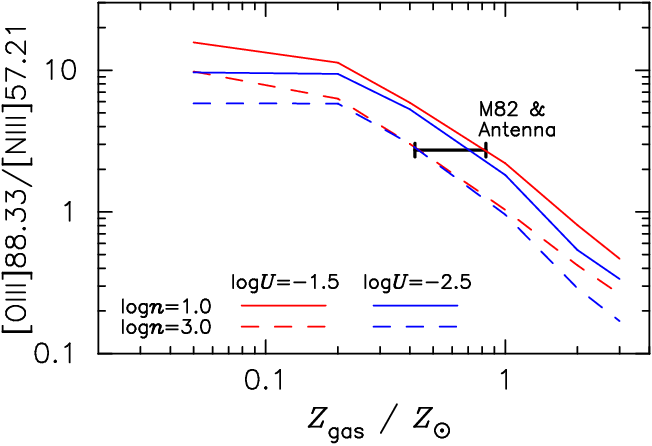}}
\caption{
   Same as Figure 3 but for the flux ratio of
   [O{\sc iii}]88.33/[N{\sc iii}]57.21. 
}
\label{fig04}
%
\rotatebox{-90}{\includegraphics[width=5.9cm]{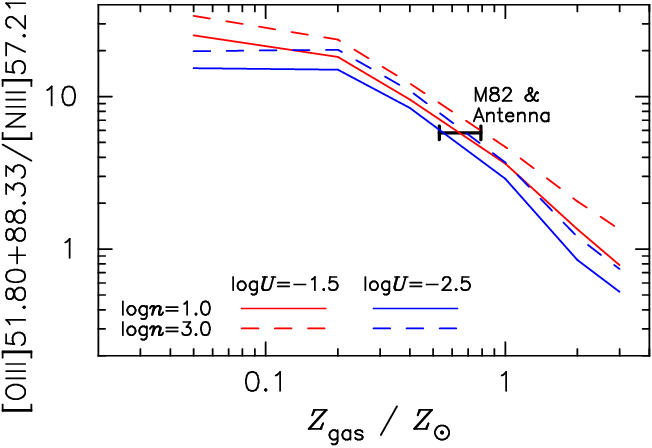}}
\caption{
   Same as Figure 3 but for the flux ratio of
   ([O{\sc iii}]51.80+[O{\sc iii}]88.33)/[N{\sc iii}]57.21. Note
   the much lower dependence on the gas density, making this ratio
   particularly suited to measure the gas metallicity.
}
\label{fig05}
\end{figure}

Liu et al. (2001) investigated N/O elemental abundance ratios of 51
Galactic planetary nebulae (PNs) and proto-planetary nebulae 
(PPNs) with [O{\sc iii}]51.80, [O{\sc iii}]88.33, and [N{\sc iii}]57.21.
They derived the N$^{2+}$/O$^{2+}$ ratio and transformed it into 
the N/O ratio. Here the two [O{\sc iii}] lines are used to correct for 
density effects. The combination of these fine-structure emission 
lines appears to be an optimal set of candidate metallicity 
diagnostics since all of their wavelength are long enough, and all 
of them are simultaneously covered by using some observational
facilities, such as the Herschel Photodetector Array Camera and 
Spectrometer (PACS) (although it requires a small redshift of 
$z \ga 0.08$ to have all lines in band). We therefore focus on this 
approach and investigate it further below.

\begin{figure}
\centering
\rotatebox{-90}{\includegraphics[width=14.6cm]{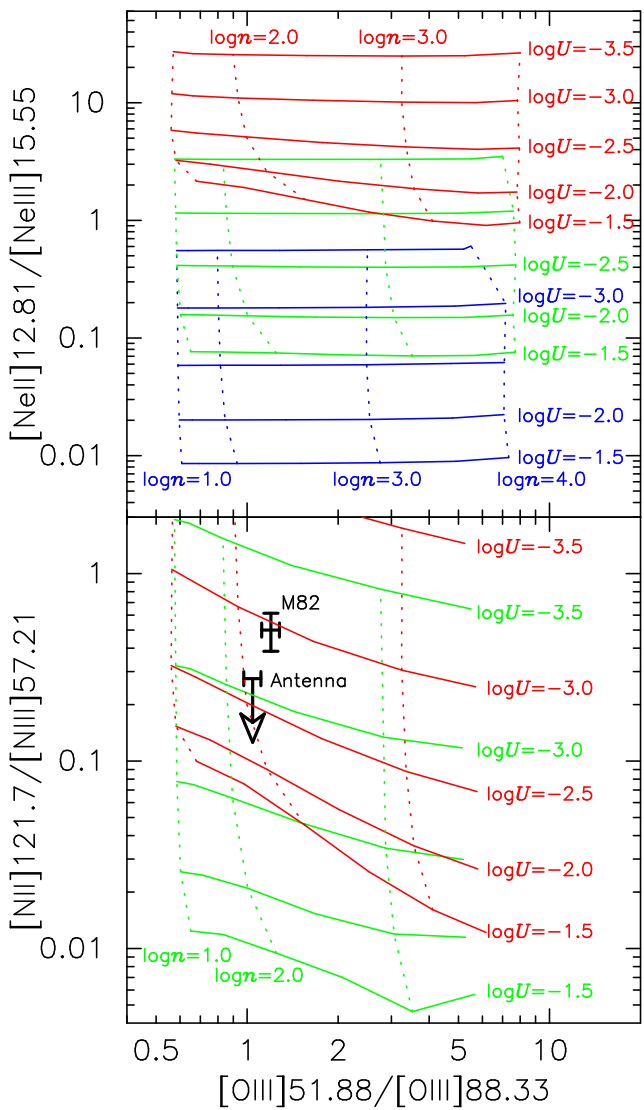}}
\caption{
  Predicted emission-line flux ratio of
  [Ne{\sc ii}]12.81/[Ne{\sc iii}]15.55 (upper) and
  [N{\sc ii}]121.7/[N{\sc iii}]57.21 (lower) as a 
  function of the density-sensitive flux 
  ratio [O{\sc iii}]51.80/[O{\sc iii}]88.33.
  Solid and dotted lines denote $U$-constant and
  $n$-constant model sequences, respectively.
  Models with $Z_{\rm gas} = $ 0.2, 1.0, and 3.0
  $Z_\odot$ are denoted by blue, green, and red lines,
  respectively. The models with $Z_{\rm gas} = 0.2 Z_\odot$ 
  or with $n_{\rm H} = 10^{4.0}$ cm$^{-3}$ are not shown 
  in the lower panel because 
  [N{\sc ii}]121.7 is extremely faint in those case. 
  Observational data for M82 and for the Antennae galaxy are 
  shown in the lower panel.
}
\label{fig06}
\end{figure}

In Figure 3, the predicted [O{\sc iii}]51.80/[N{\sc iii}]57.21 flux ratio is 
shown as a function of the gas metallicity. It is shown that this flux 
ratio has a strong metallicity dependence, especially in the high 
metallicity range ($Z_{\rm gas} > 0.2 Z_\odot$). However, this flux 
ratio also displays (smaller) dependences on the ionization 
parameter and on the gas density, which prevent to accurately infer 
the gas metallicity. Figure 4 shows the predicted flux ratios of
[O{\sc iii}]88.33/[N{\sc iii}]57.21, as a function of the gas metallicity.
Again we wee the density effects on the flux ratio. Note that models
with a lower gas density predict larger flux ratios of
[O{\sc iii}]88.33/[N{\sc iii}]57.21, that is in an opposite trend from the
case of [O{\sc iii}]51.80/[N{\sc iii}]57.21 (Figure 3).

The diagnostic accuracy improves significantly if the sum of
[O{\sc iii}]51.80 and [O{\sc iii}]88.33 is used instead of only 
[O{\sc iii}]51.80, as shown in Figure 5. The density effect on the flux 
ratio of ([O{\sc iii}]51.80+[O{\sc iii}]88.33)/[N{\sc iii}]57.21 is small 
($\la 50$\%) over the whole range of metallicities. However, there is 
still a significant dependence on the ionization parameter, especially 
at high-metallicities ($Z_{\rm gas} \ga 2 Z_\odot$), where it may 
cause variation of the ratio up to a factor of $\sim 3$. We will compare
these model predictions with observational data in \S\S4.1.

The predicted fluxes of H{\sc ii} region lines are summarized in 
Table 4, where they are normalized by the [O{\sc iii}]51.88 flux. The 
results do not change significantly when we adopt constant-density 
models instead of constant-pressure models. More specifically, the 
differences in the predicted fluxes of the emission lines given in 
Table 4 between constant-pressure models and constant-density 
models are less than 20\% in most cases.

Next we focus on how to account for the dependence of the 
ionization parameter of the flux ratios shown in Figures 3, 4, and 5.
To estimate the ionization parameter, pairs of ion species of the same 
element in a different ionization degree have been often used. We 
thus examine two flux ratios of [Ne{\sc ii}]12.81/[Ne{\sc iii}]15.55 
and [N{\sc ii}]121.7/[N{\sc iii}]57.21 as possible tools to correct for 
the ionization parameter effect. Note that also [Ne{\sc ii}]12.81 and 
[N{\sc ii}]121.7 are H{\sc ii} region lines (i.e. not from PDRs), as 
described in \S\S3.1. In Figure 6, we show these two emission-line 
flux ratios as a function of the density-sensitive flux ratio 
[O{\sc iii}]51.80/[O{\sc iii}]88.33. Clearly, both of the two flux ratios 
are strongly dependent on the ionization parameter and therefore 
they can be used to measure U and correct the metallicity diagnostic 
flux ratio ([O{\sc iii}]51.80+[O{\sc iii}]88.33)/[N{\sc iii}]57.21 for this 
dependence. These two flux ratios are also sensitive to metallicity, 
that is partly due to the metallicity-dependent SED of the ionizing 
photons (Figure 1). Therefore an iterative process is required to 
estimate metallicity of galaxies by using the diagnostic diagrams 
given in Figures 5 and 6. Which is the $U$-sensitive flux ratio more 
appropriate to correct for the ionization parameter effects on the 
metallicity determination depends on the specific case. The 
[Ne{\sc ii}]12.81 and [Ne{\sc iii}]15.55 emission lines are 
moderately strong in star-forming galaxies and have been observed 
by Spitzer in many galaxies (e.g., Dale et al. 2006), and their flux 
ratio has only a slight dependence on the gas density, as shown in 
Figure 6. However, these lines are more affected by dust extinction 
effect with respect to the metallicity diagnostic emission lines,
[O{\sc iii}]51.80, [O{\sc iii}]88.33, and [N{\sc iii}]57.21.
In contrast, [N{\sc iii}]57.21 and [N{\sc ii}]121.7 do not suffer 
significant dust extinction. However, the expected emissivity of
[N{\sc ii}]121.7 is low, which makes its measurement challenging. 
In Figure 7, we also plot the same $U$-sensitive flux ratios, 
[Ne{\sc ii}]12.81/[Ne{\sc iii}]15.55 and 
[N{\sc ii}]121.7/[N{\sc iii}]57.21 as a function of the density-sensitive 
flux ratio [N{\sc ii}]121.7/[N{\sc ii}]205.4, instead of 
[O{\sc iii}]51.80/[O{\sc iii}]88.33. Since the [N{\sc ii}]205.4 emission 
is very faint and at a very long wavelength, it is very difficult to
study with Herschel and SPICA, though its detection should be 
feasible with ALMA.

\begin{figure}
\centering
\rotatebox{-90}{\includegraphics[width=14.6cm]{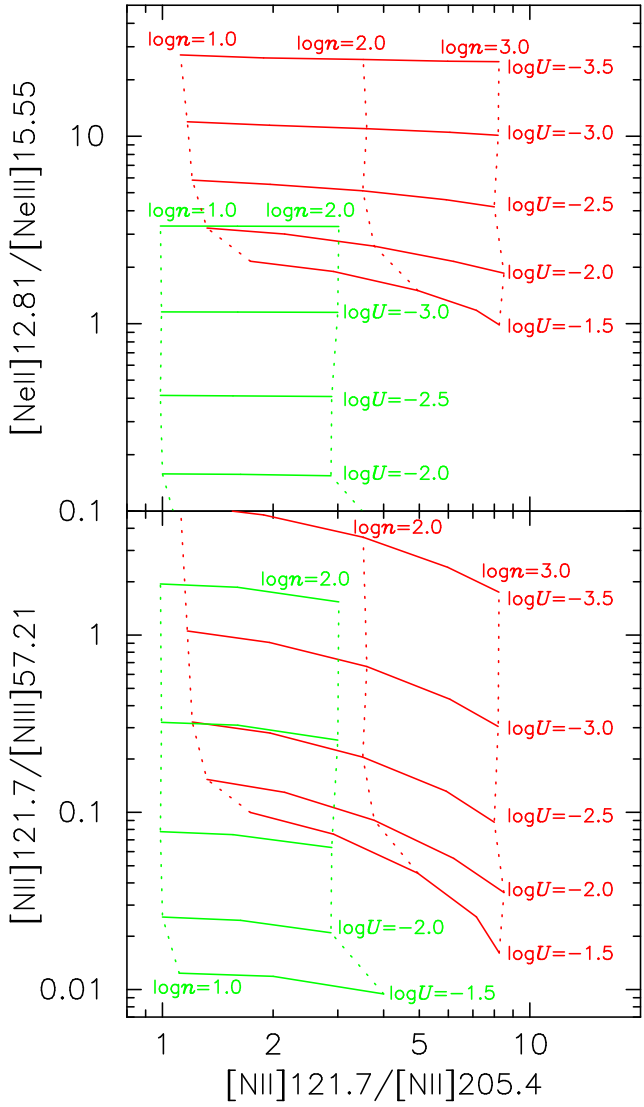}}
\caption{
  Predicted emission-line flux ratio of [Ne{\sc ii}]12.81/[Ne{\sc iii}]15.55 
  (upper) and [N{\sc ii}]121.7/[N{\sc iii}]57.21 (lower) as a function of 
  the density-sensitive flux ratio [N{\sc ii}]121.7/[N{\sc ii}]205.4.
  Solid and dotted lines denote $U$-constant and $n_{\rm H}$-constant 
  model sequences, respectively. Models with $Z_{\rm gas}$
  = 1.0 $Z_\odot$ and 3.0 $Z_\odot$ are denoted by green and red 
  lines, respectively. Only the models with $n_{\rm H} \leq 10^2$ cm$^{-3}$ 
  are plotted for the cases of $Z_{\rm gas} = 1.0 Z_\odot$, and models 
  with $n_{\rm H} \leq 10^3$ cm$^{-3}$ are plotted for the cases of 
  $Z_{\rm gas} = 3.0 Z_\odot$. This is because the
  [N{\sc ii}] lines are faint in clouds with a high density due to the
  collisional de-excitation effects.
}
\label{fig07}
\end{figure}

\subsection{Dependences on the stellar age}

\begin{figure}
\centering
\rotatebox{-90}{\includegraphics[width=10.6cm]{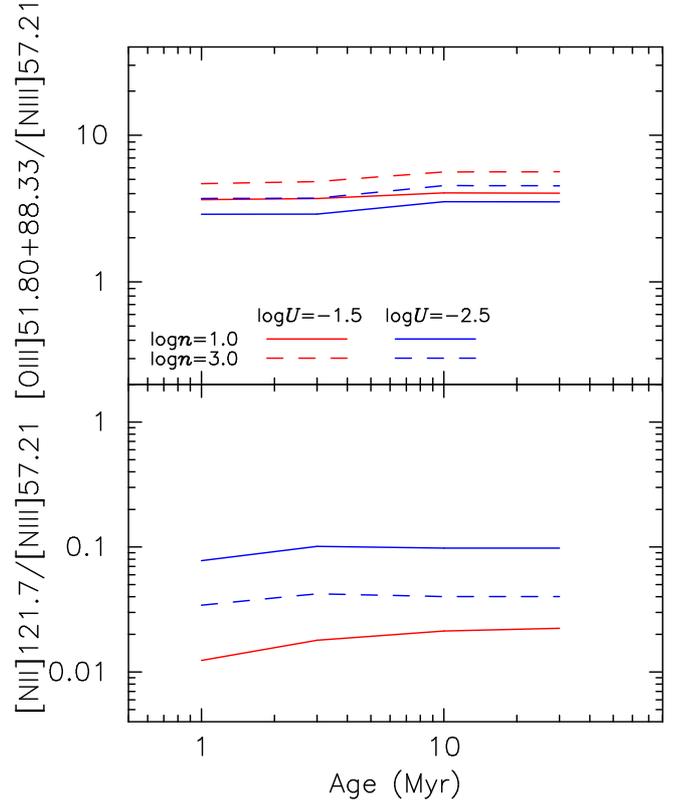}}
\caption{
  Dependence of the predicted emission-line flux ratios
  ([O{\sc iii}]51.80+[O{\sc iii}]88.33)/[N{\sc iii}]57.21
  (upper panel) and [N{\sc ii}]121.7/[N{\sc iii}]57.21
  (lower panel) on the age of the stellar population used in 
  the photoionization model calculations. Models with constant 
  star-formation history and with solar metallicity are 
  displayed. The meanings of lines are the same as in Figure 3.
  In the lower panel, models with $n_{\rm H} = 10^3$ cm$^{-3}$
  and $U = 10^{-1.5}$ are not plotted, because the flux of the
  [N{\sc ii}]121.7 is too faint for those parameters.
  The displayed ranges of emission-line flux ratios in the upper
  and lower panels are the same as those in Figures 5 and the 
  lower panel of Figure 6, respectively.
}
\label{fig08}
\end{figure}

All emission-line flux ratios shown in Figures 3--7 are
calculated by adopting constant star-formation SEDs with an age 
of 1 Myr. However, it should be verified whether the predicted 
flux ratios depend on the age of the stellar population since 
the adopted age (1 Myr) seems too young compared to the typical
age of star-forming galaxies in general. 
Cid Fernandes et al. (2003) investigated stellar populations
of nearby starburst galaxies and reported that the typical
starburst age is $\sim 10^{7.0-7.5}$ yr. More recently,
Rodr\'{\i}guez Zaur\'{\i}n et al. (2010) studied stellar 
populations of low-$z$ ULIRGs, finding similar age ranges. It 
is thus important to examine how the emission-line diagnostics
studied in this work depend on the age of the stellar 
populations used for the input SED.

We calculated the emission-line flux ratios of 
([O{\sc iii}]51.80+[O{\sc iii}]88.33)/[N{\sc iii}]57.21 and
[N{\sc ii}]121.7/[N{\sc iii}]57.21 by adopting the input SED
of stellar populations with ages of 3, 10, and 30 Myr
in addition to our default value (1 Myr).
Figure 8 shows the results of the age dependences of those
two diagnostics. The flux ratio of
([O{\sc iii}]51.80+[O{\sc iii}]88.33)/[N{\sc iii}]57.21 
changes by less than 30\% in the range $1-30$ Myr, that is
negligible with respect to the metallicity dependence of
this flux ratio shown in Figure 5. The age dependence is
not significant also for the flux ratio 
[N{\sc ii}]121.7/[N{\sc iii}]57.21.

These results are due to the fact that the SED shape 
significantly changes as a function of the stellar-population 
age (for the range of $1-30$ Myr) only at 
$\lambda > 3000{\rm \AA}$ (see Figure 8 in Leitherer et al. 
1999), when the constant star-formation history is assumed.
The SED at shorter wavelengths does not significantly depend
on the starburst age, because in that wavelength range it
is dominated by emission from massive stars.
Since the ionization structure in H{\sc ii} regions is mostly 
determined by ultraviolet photons (whose SED is not sensitive 
to the stellar-population age), the flux ratios of 
H{\sc ii}-region lines are consequently insensitive to the
starburst age.
These situation may be completely different for PDR lines
such as [C{\sc ii}]157.6, but the age dependence of such
PDR lines is beyond the scope of this work.
Note that the constant star-formation history is a better
assumption than the single-burst model for star-forming 
galaxies when relatively old ages ($\sim 30$ Mye) are 
assumed, because stellar populations with an age of 
$\sim 30$ Myr in the single-burst model do not produce 
ionizing photons and thus we do not expect detectable
emission-line fluxes from such systems. It can be therefore 
concluded that our analysis on metallicity (and other) 
diagnostics does not significantly depend on the age of 
stellar populations.

\section{Discussion}

\subsection{Comparison with observational data}

Here we compare our calculation results on long-wavelength
($\lambda > 50 \mu$m) diagnostic fine structure lines with
existing observational data. The only previous instrument that 
measured the diagnostic fine-structure lines in 
star-forming galaxies is LWS onboard $ISO$. 
Specifically we focus on two galaxies, M82 and the
Antennae galaxy, since many infrared fine-structure lines 
are measured for them (Fischer et al. 1996; Colbert 
et al. 1999). The measured fluxes are summarized in 
Table 5. Interestingly, the measured
flux ratio of [O{\sc iii}]51.80/[O{\sc iii}]88.33
is quite similar between those two galaxies;
1.2 and 1.0 for M82 and the Antennae. These flux 
ratios suggest $n_{\rm H} \sim 10^2$ cm$^{-3}$ for 
$U \la 10^{-2}$ (Figure 5). The metallicity
diagnostic flux ratio of
([O{\sc iii}]51.80+[O{\sc iii}]88.33)/[N{\sc iii}]57.21
is also similar between M82 and the Antennae,
5.6 and 6.0 respectively. Taking the inferred gas
density also into account, these flux ratios 
correspond to $Z \sim 0.5-0.7 Z_\odot$, i.e.,
slightly sub-solar metallicities.
The observed flux ratio of 
[N{\sc ii}]121.7/[N{\sc iii}]57.21 in M82 is measured 
to be 0.5, which, combined with the our initial metallicity estimation,
gives $\rm \log U = -3.5$. Only an upper limit of
[N{\sc ii}]121.7/[N{\sc iii}]57.21$< 0.28$ 
(3 $\sigma$) is reported for the Antennae galaxy, providing an upper
limit on $\rm \log U$ of about $-3$. Although the 
constraint on the ionization parameter is not strong, 
the uncertainty on the ionization parameter does not 
affect significantly the metallicity estimation.

Although both M82 and the Antennae are well-studied
nearby galaxies, it is not so straightforward to
compare the inferred gas metallicity based on the
fine-structure emission-line flux ratios with that
inferred from other metallicity diagnostics, 
because the aperture size of LWS (the beam-size is
$\sim 80$ arcsec in FWHM) is very large and difficult to compare
with other measurements.
Origlia et al. (2004) measured the nuclear gaseous
and stellar metallicities by using X-ray and near-infrared
spectroscopic observations, and they found that
both the metallicities are close to or slightly
less than the solar metallicity in M82. Taking the 
metallicity gradient into account, these metallicities (based also
on diagnostics little affected by dust extinction) are
consistent with the metallicity inferred through
the far-IR fine-structure diagnostics. The situations is
similar also for the Antennae galaxy, since its
gaseous and stellar metallicities are close to or
slightly less than the solar metallicity (Bastian
et al. 2006, 2009). Although the comparisons discussed above
do not provide a tight test for our new metallicity diagnostics
based on far-IR fine structure lines, do at least
suggest a broad consistency with other 
metallicity diagnostics.

\subsection{Observational feasibility with next-generation instruments}

\begin{figure}
\centering
\rotatebox{0}{\includegraphics[width=9.6cm]{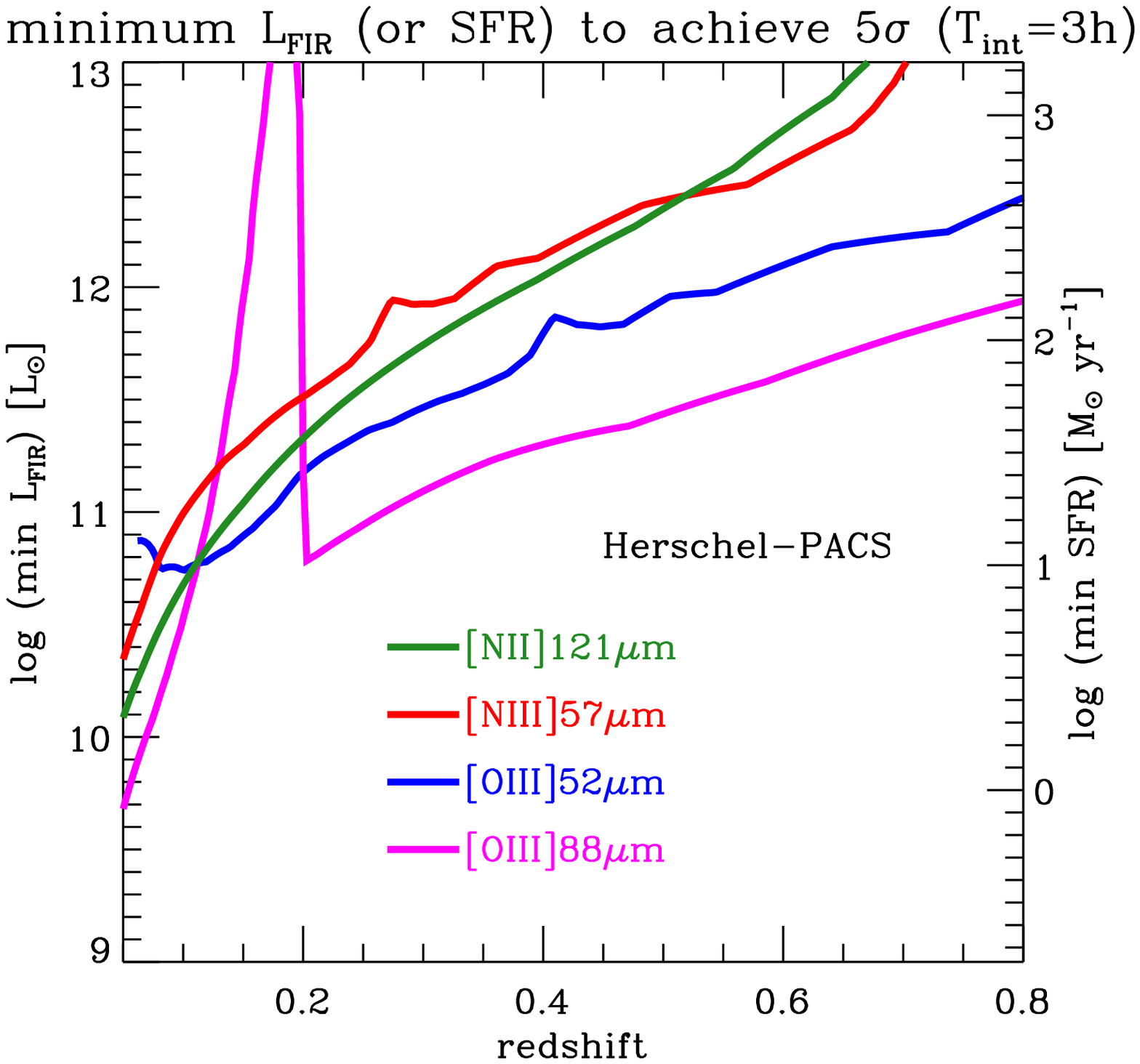}}
\caption{
  Detectability with PACS on Herschel
  of some diagnostic far-IR fine-structure
  lines discussed in this paper, which can be used to measure
  the gas metallicity, ionization parameter and density in galaxies,
  as a function of redshift.
  The left hand scale gives minimum $L_{\rm FIR}$
  to achieve a 5$\sigma$ detection of each line
  with an exposure time of 3 hours. The right hand axis shows
  the minimum star formation rate for the line detectability
  corresponding to the minimum $L_{\rm FIR}$
  based on the relation given in Kennicutt (1999).
  Green, red, blue, and magenta lines denote [N{\sc ii}]121.7,
  [N{\sc iii}]57.21, [O{\sc iii}]51.80, and [O{\sc iii}]88.33,
  respectively.
}
\label{fig09}
%
\rotatebox{0}{\includegraphics[width=9.6cm]{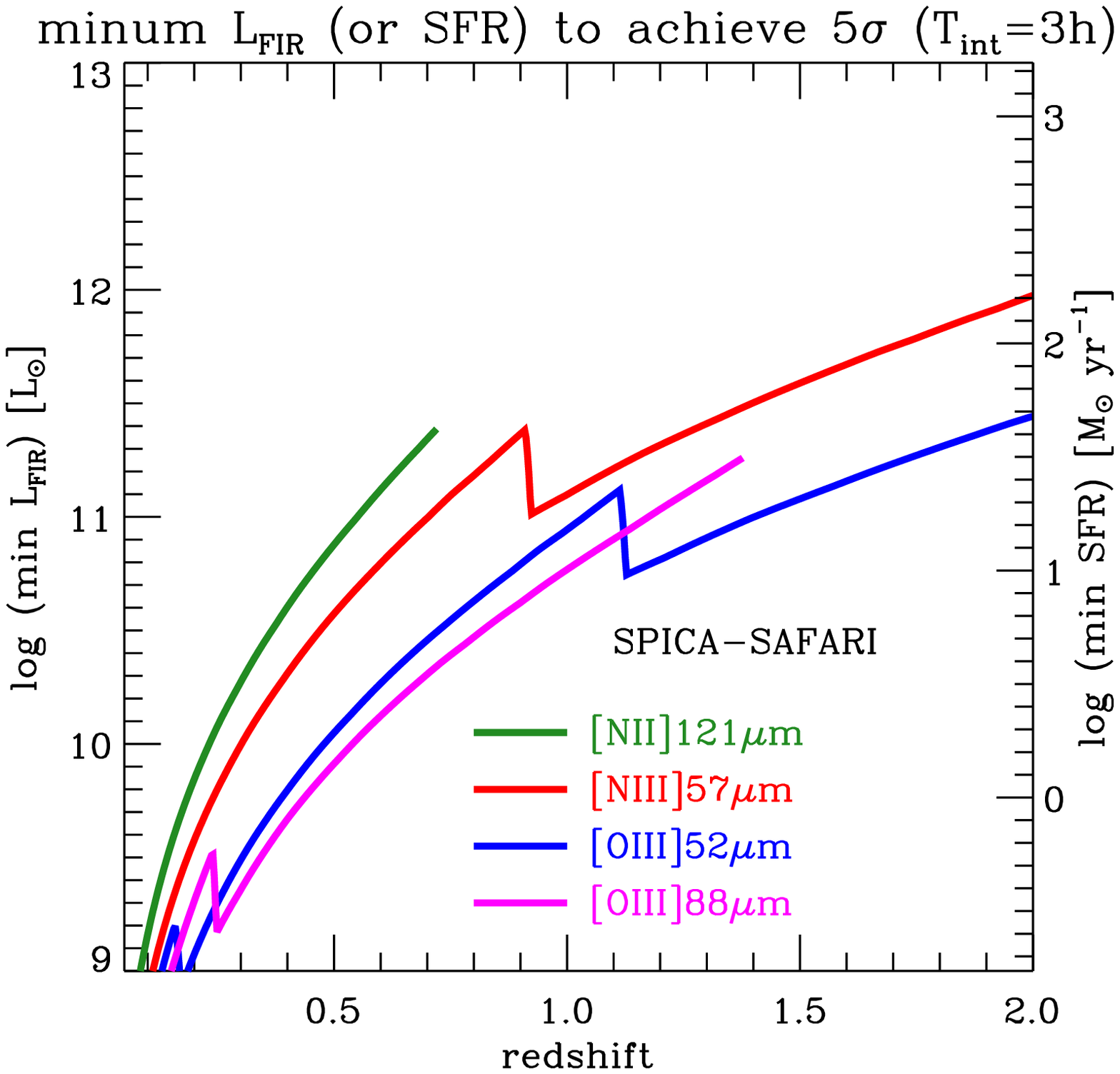}}
\caption{
  Same as Figure 9 but for the detectability with 
  SAFARI boarded on SPICA.
}
\label{fig10}
\end{figure}

We finally examine the detectability of the diagnostic 
fine-structure emission lines discussed here with the
new-generation instruments, PACS on Herschel (Poglitsch et al. 
2010) and SAFARI (Swinyard 2008; Swinyard et al. 2009) the 
spectrometer planned for SPICA (a 3m-class cooled telescope in 
space, Nakagawa 2009). Figure 9 shows the detectability of 
[N{\sc ii}]121.7, [N{\sc iii}]57.21, [O{\sc iii}]51.80, and 
[O{\sc iii}]88.33 with PACS. Specifically the figure shows the 
minimum far-infrared luminosity of galaxies to have each far-IR 
line detected (S/N=5) with a 3 hour exposure, as a function of 
redshift. Here we make the simplifying assumption that the 
luminosity of these emission lines scales roughly with 
far-infrared luminosity and keeping a ratio similar, on average, 
to what observed in some nearby galaxies where all of these 
transitions were observed with past ISO observations; more 
specifically,
$L_{\rm [OIII]88.33}/L_{\rm FIR} \sim 1.8 \times 10^{-3}$,
$L_{\rm [OIII]51.80}/L_{\rm FIR} \sim 1.8 \times 10^{-3}$,
$L_{\rm [NIII]57.21}/L_{\rm FIR} \sim 6.0 \times 10^{-4}$, and
$L_{\rm [NII]121.7}/L_{\rm FIR}  \sim 2.7 \times 10^{-4}$
(Fischer et al. 1996; Colbert et al. 1999). The scale on the 
right hand side gives the minimum SFR corresponding the minimum 
$\rm L_{FIR}$, by using the SFR--$\rm L_{FIR}$ relationship 
given in Kennicutt (1999). Figure 9 suggests that we can detect 
all the fine-structure emission lines needed to infer the 
metallicity, in luminous infrared galaxies 
($\rm L_{FIR}\sim 10^{11}-10^{12} L_{\odot}$) at redshifts up 
to $z \sim 0.4$. The detection to somewhat higher redshift is 
restricted to ULIRGs ($\rm L_{FIR}>
10^{12} ~L_{\odot}$). On the other hand, SAFARI boarded on 
SPICA will drastically expand the detectability, as shown in 
Figure 10. By using SAFARI, the diagnostic fine-structure lines
will be detectable for star-forming galaxies at $z \sim 1$ or 
at even higher redshift. In Figure 10 we adopt the SAFARI 
sensitivity given by Swinyard (2008) with a scaling of 
sensitivity for a conservative 3.0m aperture. Actually the 
redshift limit may be limited not by the instrument sensitivity, 
but by the wavelength coverage, although the instrument design 
of SAFARI has not yet completely fixed.

\section{Summary}

Gas metallicity diagnostics with infrared fine-structure emission lines
are investigated in this paper, based on detail photoionization model
calculations. The main results are as follows:
\begin{itemize}
  \item Some fine-structure lines arise both in H{\sc ii} regions and PDRs
            and thus only emission lines arising within H{\sc ii} regions (not
            in PDRs) should be used as good metallicity diagnostics.
  \item Emission-line flux ratios that consist of [O{\sc iii}]51.80, 
            [N{\sc iii}]57.21, and [O{\sc iii}]88.33 provide good metallicity 
            diagnostics; especially the flux ratio of
            ([O{\sc iii}]51.80+[O{\sc iii}]88.33)/[N{\sc iii}]57.21 is sensitive to
            the gas metallicity with a small dependence on the gas density.
  \item To correct for the effects of the ionization parameter on the
            infrared metallicity diagnostics, flux ratios of 
            [Ne{\sc ii}]12.81/[Ne{\sc iii}]15.55 and 
            [N{\sc ii}]121.7/[N{\sc iii}]57.21 are useful.
  \item The detection of those infrared fine-structure lines is feasible,
            up to $z \sim 0.4$ with using Herschel/PACS and even up to
            $z \sim 1$ or more with using SPICA/SAFARI.
\end{itemize}
Since the infrared fine-structure emission lines are not affected by 
the dust extinction, the gas-metallicity diagnostics presented in this 
paper are useful to assess the chemical properties in dust-obscured 
populations such as infrared galaxies, that are very important in
terms of the galaxy evolution. Some of the infrared fine-structure 
emission lines from high-$z$ galaxies will be apparently exciting 
targets also for ALMA.

\begin{acknowledgements}
  We thank Walmsley Malcolm, Luigi Spinoglio, Yasuhiro Shioya, 
  and the anonymous referee for useful comments and suggestions. 
  TN acknowledges financial supports through the Research 
  Promotion Award of Ehime University and the Kurata Grant from 
  the Kurata Memorial Hitachi Science and Technology Foundation.
  AM and RM acknowledge support from the Italian National Institute
  for Astrophysics through a PRIN/INAF 2008 grant.
\end{acknowledgements}


\begin{table*}
\centering
\caption{Fractions of the H{\sc ii} region contribution
         to total emission-line fluxes
         in low-metallicity cases ($Z_{\rm gas} = 0.2 Z_\odot$)}
\label{table:02}
\begin{tabular}{l c c c c}
\hline\hline
\noalign{\smallskip}

&
$n_{\rm H} = 10^{1.0}$ cm$^{-3}$ &
$n_{\rm H} = 10^{1.0}$ cm$^{-3}$ &
$n_{\rm H} = 10^{3.0}$ cm$^{-3}$ &
$n_{\rm H} = 10^{3.0}$ cm$^{-3}$ \\
Line &
$U = 10^{-2.5}$ &  
$U = 10^{-1.5}$ &  
$U = 10^{-2.5}$ &  
$U = 10^{-1.5}$ \\ 

\noalign{\smallskip}
\hline 
\noalign{\smallskip}

{}[Ne {\sc ii}]  12.81 & 
    1.00 & 1.00 & 1.00 & 1.00 \\
{}[Ne {\sc iii}] 15.55 & 
    1.00 & 1.00 & 1.00 & 1.00 \\
{}[S {\sc iii}]  18.67 & 
    1.00 & 1.00 & 1.00 & 1.00 \\
{}[S {\sc iii}]  33.47 & 
    1.00 & 1.00 & 1.00 & 1.00 \\
{}[Si {\sc ii}]  34.81 & 
    0.99 & 0.80 & 0.23 & 0.05 \\
{}[Ne {\sc iii}] 36.01 & 
    1.00 & 1.00 & 1.00 & 1.00 \\
{}[O {\sc iii}]  51.80 & 
    1.00 & 1.00 & 1.00 & 1.00 \\
{}[N {\sc iii}]  57.21 & 
    1.00 & 1.00 & 1.00 & 1.00 \\ 
{}[O {\sc i}]    63.17 & 
    0.70 & 0.05 & 0.02 &$<$0.01\\
{}[O {\sc iii}]  88.33 & 
    1.00 & 1.00 & 1.00 & 1.00 \\
{}[N {\sc ii}]   121.7 & 
    1.00 & 1.00 & 1.00 & 1.00 \\
{}[O {\sc i}]    145.5 & 
    0.93 & 0.21 & 0.05 &$<$0.01\\
{}[C {\sc ii}]   157.6 & 
    0.33 & 0.03 & 0.03 &$<$0.01\\
{}[N {\sc ii}]   205.4 & 
    1.00 & 1.00 & 1.00 & 1.00 \\

\noalign{\smallskip}
\hline
\end{tabular}
\end{table*}

\begin{table*}
\centering
\caption{Fractions of the H{\sc ii} region contribution
         to total emission-line fluxes
         in high-metallicity cases ($Z_{\rm gas} = 3 Z_\odot$)}
\label{table:03}
\begin{tabular}{l c c c c}
\hline\hline
\noalign{\smallskip}

&
$n_{\rm H} = 10^{1.0}$ cm$^{-3}$ &
$n_{\rm H} = 10^{1.0}$ cm$^{-3}$ &
$n_{\rm H} = 10^{3.0}$ cm$^{-3}$ &
$n_{\rm H} = 10^{3.0}$ cm$^{-3}$ \\
Line &
$U = 10^{-2.5}$ &  
$U = 10^{-1.5}$ &  
$U = 10^{-2.5}$ &  
$U = 10^{-1.5}$ \\ 

\noalign{\smallskip}
\hline 
\noalign{\smallskip}

{}[Ne {\sc ii}]  12.81 & 
    1.00 & 1.00 & 1.00 & 1.00 \\
{}[Ne {\sc iii}] 15.55 & 
    1.00 & 1.00 & 1.00 & 1.00 \\
{}[S {\sc iii}]  18.67 & 
    1.00 & 1.00 & 1.00 & 1.00 \\
{}[S {\sc iii}]  33.47 & 
    1.00 & 1.00 & 1.00 & 1.00 \\
{}[Si {\sc ii}]  34.81 & 
    1.00 & 0.96 & 0.61 & 0.37 \\ 
{}[Ne {\sc iii}] 36.01 & 
    1.00 & 1.00 & 1.00 & 1.00 \\
{}[O {\sc iii}]  51.80 & 
    1.00 & 1.00 & 1.00 & 1.00 \\
{}[N {\sc iii}]  57.21 & 
    1.00 & 1.00 & 1.00 & 1.00 \\
{}[O {\sc i}]    63.17 & 
    0.57 & 0.06 & 0.03 & 0.02 \\
{}[O {\sc iii}]  88.33 & 
    1.00 & 1.00 & 1.00 & 1.00 \\
{}[N {\sc ii}]   121.7 & 
    1.00 & 1.00 & 1.00 & 1.00 \\
{}[O {\sc i}]    145.5 & 
    0.82 & 0.12 & 0.05 & 0.02 \\
{}[C {\sc ii}]   157.6 & 
    0.75 & 0.34 & 0.15 & 0.11 \\
{}[N {\sc ii}]   205.4 & 
    1.00 & 1.00 & 1.00 & 1.00 \\

\noalign{\smallskip}
\hline
\end{tabular}
\end{table*}


\begin{table*}
\centering
\caption{Predicted emission-line fluxes by photoionization models}
\label{table:04}
\begin{tabular}{c c c c c c c c c c}
\hline\hline
\noalign{\smallskip}

\multicolumn{3}{c}{Model Parameter} &
\multicolumn{7}{c}{Predicted Fluxes Normalized by $F$([O{\sc iii}]51.88)} \\

log $n_{\rm H}$$^\mathrm{a}$ & 
log $U$                      & 
$Z_{\rm gas}/Z_\odot$        & 
[Ne{\sc ii}]12.81            & 
[Ne{\sc iii}]15.55           &
[O{\sc iii}]51.88            &
[N{\sc iii}]57.21            &
[O{\sc iii}]88.33            &
[N{\sc ii}]121.7             &
[N{\sc ii}]205.4            \\

\noalign{\smallskip}
\hline 
\noalign{\smallskip}

1.0 & --1.5 & 0.05 & 0.004 & 0.809 & 1.000 & 0.106 & 1.664&$\leq$0.001&$\leq$0.001\\
1.0 & --1.5 & 0.20 & 0.006 & 0.789 & 1.000 & 0.145 & 1.646&$\leq$0.001&$\leq$0.001\\
1.0 & --1.5 & 0.40 & 0.011 & 0.769 & 1.000 & 0.274 & 1.617 & 0.002 & 0.002 \\
1.0 & --1.5 & 1.00 & 0.054 & 0.699 & 1.000 & 0.698 & 1.536 & 0.009 & 0.008 \\
1.0 & --1.5 & 2.00 & 0.394 & 0.396 & 1.000 & 1.845 & 1.488 & 0.078 & 0.055 \\
1.0 & --1.5 & 3.00 & 0.618 & 0.287 & 1.000 & 3.147 & 1.471 & 0.314 & 0.181 \\
1.0 & --2.5 & 0.05 & 0.031 & 0.928 & 1.000 & 0.171 & 1.691 & 0.006 & 0.006 \\
1.0 & --2.5 & 0.20 & 0.053 & 0.916 & 1.000 & 0.179 & 1.693 & 0.007 & 0.007 \\
1.0 & --2.5 & 0.40 & 0.089 & 0.899 & 1.000 & 0.320 & 1.698 & 0.014 & 0.015 \\
1.0 & --2.5 & 1.00 & 0.322 & 0.778 & 1.000 & 0.937 & 1.708 & 0.073 & 0.074 \\
1.0 & --2.5 & 2.00 & 1.382 & 0.409 & 1.000 & 3.229 & 1.737 & 0.598 & 0.543 \\
1.0 & --2.5 & 3.00 & 1.916 & 0.328 & 1.000 & 5.308 & 1.783 & 1.714 & 1.423 \\
3.0 & --1.5 & 0.05 & 0.004 & 0.942 & 1.000 & 0.042 & 0.408&$\leq$0.001&$\leq$0.001\\
3.0 & --1.5 & 0.20 & 0.008 & 0.984 & 1.000 & 0.058 & 0.362&$\leq$0.001&$\leq$0.001\\
3.0 & --1.5 & 0.40 & 0.015 & 1.039 & 1.000 & 0.109 & 0.329&$\leq$0.001&$\leq$0.001\\
3.0 & --1.5 & 1.00 & 0.080 & 1.132 & 1.000 & 0.275 & 0.284&$\leq$0.001&$\leq$0.001\\
3.0 & --1.5 & 2.00 & 0.529 & 0.825 & 1.000 & 0.612 & 0.257 & 0.006    &$\leq$0.001\\
3.0 & --1.5 & 3.00 & 0.767 & 0.787 & 1.000 & 0.924 & 0.243 & 0.015 & 0.002 \\
3.0 & --2.5 & 0.05 & 0.036 & 1.057 & 1.000 & 0.068 & 0.417 &$\leq$0.001&$\leq$0.001\\
3.0 & --2.5 & 0.20 & 0.062 & 1.061 & 1.000 & 0.071 & 0.401 &$\leq$0.001&$\leq$0.001\\
3.0 & --2.5 & 0.40 & 0.106 & 1.067 & 1.000 & 0.127 & 0.383 & 0.003     &$\leq$0.001\\
3.0 & --2.5 & 1.00 & 0.395 & 0.986 & 1.000 & 0.364 & 0.348 & 0.013 & 0.002 \\
3.0 & --2.5 & 2.00 & 1.607 & 0.553 & 1.000 & 1.094 & 0.316 & 0.069 & 0.009 \\
3.0 & --2.5 & 3.00 & 2.245 & 0.534 & 1.000 & 1.757 & 0.298 & 0.155 & 0.019 \\

\noalign{\smallskip}
\hline
\end{tabular}
\begin{list}{}{}
\item[$^\mathrm{a}$]
  Total hydrogen density at the illuminated surface, in units of
  cm$^{-3}$.
\end{list}
\end{table*}

\begin{table*}
\centering
\caption{Measured fluxes of infrared fine-structure lines
         in M82 and the Antennae galaxy$^\mathrm{a}$}
\label{table:05}
\begin{tabular}{l c c}
\hline\hline
\noalign{\smallskip}

line                 &
M82$^\mathrm{b}$     &
Antenna$^\mathrm{c}$ \\

\noalign{\smallskip}
\hline 
\noalign{\smallskip}

{}[O{\sc iii}]51.80 
   & $10.3 \pm 0.5  \times 10^{-18}$
   & $ 4.9 \pm 0.29 \times 10^{-19}$ \\
{}[N{\sc iii}]57.21
   & $ 3.4 \pm 0.5  \times 10^{-18}$
   & $ 1.6 \pm 0.058\times 10^{-19}$ \\
{}[O{\sc iii}]88.33
   & $ 8.6 \pm 0.4  \times 10^{-18}$
   & $ 4.7 \pm 0.13 \times 10^{-19}$ \\
{}[N{\sc ii}]121.7
   & $ 1.7 \pm 0.3  \times 10^{-18}$
   & $ <0.44 \times 10^{-19}$ (3$\sigma$) \\

\noalign{\smallskip}
\hline
\end{tabular}
\begin{list}{}{}
\item[$^\mathrm{a}$]
  Line fluxes are given in units of W cm$^{-2}$.
\item[$^\mathrm{b}$]
  Data taken from Colbert et al. (1999).
\item[$^\mathrm{c}$]
  Data taken from Fischer et al. (1996). 
\end{list}
\end{table*}

\end{document}